\documentclass[prb,reprint,superscriptaddress]{revtex4-1}

 \pdfoutput=1

\usepackage{amsmath,amssymb}
\usepackage{graphicx,import}
\usepackage{hyperref}
\usepackage{natbib}

\begin{document}

\title{Spin liquid phases of Mott insulating ultracold bosons}

\author{Todd C. Rutkowski}
\email{trutkow1@binghamton.edu}
\affiliation{Dept. of Physics, Applied Physics and Astronomy, Binghamton University, Vestal, NY 13850.}
\affiliation{Dept. of Physics, Cornell University, Ithaca, NY, 14853}

\author{Michael J. Lawler}
\affiliation{Dept. of Physics, Applied Physics and Astronomy, Binghamton University, Vestal, NY 13850.}
\affiliation{Dept. of Physics, Cornell University, Ithaca, NY, 14853}

\date{\today}

\begin{abstract}
Mott insulating ultracold gases posses a unique whole-atom exchange interaction which enables large quantum fluctuations between the Zeeman sublevels of each atom.  By strengthening this interaction---either through the use of large-spin atoms, or by tuning the particle-particle interactions via optical Feshbach resonance---one may enhance fluctuations and facilitate the appearance of the long sought-after quantum spin liquid phase---all in the highly tunable environment of cold atoms.  To illustrate the relationship between the spin magnitude, interaction strength, and resulting magnetic phases, we present and solve a mean field theory for bosons optically confined to the one particle-per-site Mott state, using both analytic and numerical methods.  We find on a square lattice with bosons of hyperfine spin $f>2$, that making the repulsive s-wave scattering length through the singlet channel small---relative to the higher-order scattering channels---accesses a short-range resonating valence bond (s-RVB) spin liquid phase.
\end{abstract}

\pacs{2,34}

\maketitle 

\section{Introduction}
Quantum spin liquids---insulating magnetic phases which remain disordered down to absolute zero temperature---have attracted great interest in the nearly 30 years since Anderson suggested an intimate relation between the cuprate superconductors and the resonating valence bond state. \cite{Anderson1987,Lee2006}  While theorists now have a good understanding of the topological orders and fractionalized excitations that characterize these systems,\cite{Wen2002,Senthil2001} the experimental realization of the spin liquid phase has remained a challenge, and despite success in recent years\cite{Han2012,Fennell2014,Isono2014} the pool of spin liquid candidate materials remains small.  The primary difficulty lies in finding systems with sufficiently large spin fluctuations, and to achieve this in the solid-state---where electron exchange mediates the spin-spin interactions---one must restrict the search to low-dimensional, geometrically-frustrated, spin-$1/2$ antiferromagnets.\cite{Balents2010}  We believe however, that by fundamentally broadening our search to include other novel systems, we may bypass these restrictions and expedite the study of this long sought-after phase.

\begin{figure}
\includegraphics[scale=0.32]{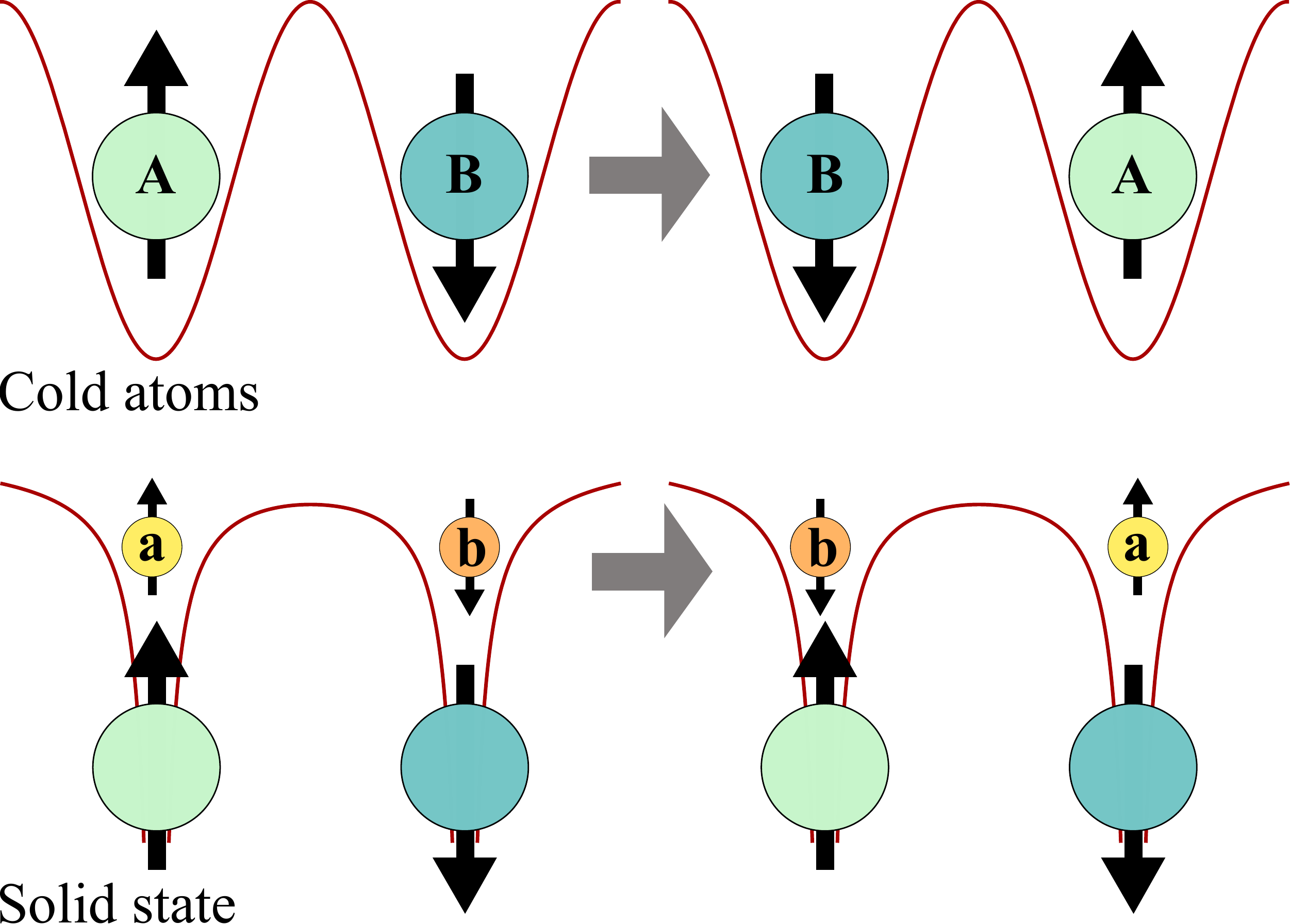}
\caption{\label{fig:Hopping} \emph{Comparison of superexchange mechanisms.}  For atomic gases optically confined to the Mott state (top), superexchange of whole spin-$f$ atoms leads to magnetic fluctuations on the order of $2f$.  In the solid state by contrast (bottom), electron superexchange restricts fluctuations to order 1, which may be small compared to the total spin on each site.  Therefore, large-spin cold atoms could potentially realize exotic fluctuation-driven states not accessible with solid state systems.  Figure adapted from Ref. \onlinecite{Wu2010}.}
\end{figure}

In particular, Mott insulating ultracold atoms may provide an alternate route to the experimental realization of spin liquids.  The spin degree of freedom remains unfrozen in these optically confined systems,\cite{Stamper-Kurn1998} and the virtual exchange of a whole atom mediates the low-energy spin-spin interaction, as illustrated by Fig. \ref{fig:Hopping}.  Counter-intuitively, whole-atom exchange produces fluctuations that increase with the atomic hyperfine spin $f$,\cite{Wu2010} in dramatic contrast to the solid-state, where large spins actually suppress the effect of fluctuations.  This peculiar behavior may cause large-$f$ Mott insulators to exhibit many exotic phases,\cite{Demler2002,Wu2006,Gorshkov2010,Szirmai2011,Cazalilla2014} including atomic spin liquids.  While efforts have focused thus far on SU($N$) symmetric alkali-earth-metal atoms,\cite{Hermele2009,Sinkovicz2013} whole-atom exchange should induce large fluctuations generically, potentially bringing the spin liquid phase to life in a wide variety of cold atomic systems.

In this paper, we perform a mean field theory for spin-$f$ bosons on a square optical lattice, tightly confined to the Mott insulating state.  We choose to include only the fundamental low-energy scattering interactions; described by a set of s-wave scattering lengths $a_F$,\cite{Ho1998} and tuned via microwave and optical Feshbach resonances.\cite{Julienne2005,Enomoto2008,Yan2013,Papoular2010}  We then show that our model supports the existence of a short-range resonating valence bond ground state, for certain values of $a_F$ and $f$.  At present, we omit both anisotropic dipole-dipole interactions and gauge fluctuations of the mean field, though we briefly discuss their effects in later sections of this paper.  Despite these omissions in our model, the results indicate that whole-atom exchange may melt magnetic order and stabilize spin liquid phases in a much broader class of systems than currently under investigation.

\section{General spin-$f$ model}

We first write a model that captures the physics of whole-atom exchange, while fluidly describing bosons of different spin.  To do this we begin with the spin-$f$ Bose-Hubbard model,\cite{Fisher1989,Jaksch1998}
\begin{multline}\label{eq:BoseHubbard}
\hat{H} = - t \sum_{<i,j>} \sum_{m=-f}^{f}(\hat{b}^\dagger_{i,m} \hat{b}_{j,m} + \text{h.c.})\\ + U \sum_i \sum_{F=0,2,\ldots}^{2f}a_F \hat{P}^F_{i},
\end{multline}
where $i$ ranges over all $N$ lattice sites, and $<i,j>$ denotes a sum over all distinct nearest-neighbor pairs.  We have written the kinetic contribution---parameterized by $t$---in terms of $\hat{b}^\dagger_{i,m}$($\hat{b}_{i,m}$) operators, which create(annihilate) a spin-$f$ boson at site $i$ with magnetic quantum number $m$, while we have expressed the on-site interaction---parameterized by $U$---in terms of projection operators $\hat{P}^{F}_{i} =\sum_{M = -F,\ldots,F} |F,M\rangle \langle F,M|$, which project the two-particle states of site $i$ into the subspace with total angular momentum $F$.  As mentioned previously, $a_F$ denotes the s-wave scattering length through the total angular momentum channel $F$.   Throughout this paper sums over $F$ and $M$ imply a range of $F = 0,2,\ldots,2f$ and $M = -F,\ldots,F$, where the absence of the odd-F scattering lengths ensures proper particle statistics.\cite{Kawaguchi2012}  Furthermore, we only consider monotonically increasing repulsive interactions, such that $a_F>0$ for all $F$, and $a_{F'}\geq a_{F}$ for $F'>F$.  In this regime, antiferromagnetic interactions dominate---a necessary condition for the non-trivial magnetic ordering that we seek.

To move to the deep Mott limit, one quenches the kinetic energy relative to the on-site repulsion\cite{Greiner2002} ($t << U$), allowing a perturbative expansion of Eq. (\ref{eq:BoseHubbard}) to second order in $t/U$.  By doing so, we obtain a spin-spin interaction in the one particle-per-site Hilbert space, which in agreement with Ref. \onlinecite{Imambekov2003,Eckert2007} yields
\begin{equation}\label{eq:ProjectionHamiltonian}
\hat{H} = - J \sum_{<i,j>} \sum_{F} \frac{1}{a_F}\hat{P}^F_{i,j},
\end{equation}
where the exchange energy is set by $J = 4t^2/U$, and the projection operator $\hat{P}^{F}_{i,j}$ now projects two sites $i$ and $j$ into total angular momentum state $F$.  The natural decoupling of the interaction into total angular momentum channels, each parametrized by a scattering length $a_F$, arises from the rotational symmetry of the low-energy interaction, which conserves the total angular momentum of two bosons during a collision\cite{Ho1998}.

At this point, one commonly re-expresses the $\hat{P}^{F}_{i,j}$ operators of Eq. (\ref{eq:ProjectionHamiltonian}) in terms of operators which possess a more direct physical interpretation, such as a polynomial in the Heisenberg coupling $\hat{\mathbf{S}}_i \cdot \hat{\mathbf{S}}_j$, or with tensor operators of increasing rank.  These methods do not move fluidly from one spin $f$ to another however, as one must continually define new operators upon increasing the spin. Although schemes have been developed to simplify such descriptions,\cite{Barnett2006,Barnett2007} we instead elect to return our Hamiltonian to second quantization.  While doing so simplifies study for generic $f$, it implicitly enlarges our Hilbert space to include multiply-occupied sites.  Therefore, to maintain equivalence between the two Hilbert spaces we must impose a one particle-per-site constraint on average.\cite{Wen1991}  With these considerations in mind, we write our Hamiltonian in second quantization as
\begin{equation}\label{eq:FullHamiltonian}
\hat{H} = - J \sum_{<i,j>} \sum_{F,M} \frac{1}{a_F}\hat{A}^{FM \dagger}_{i,j} \hat{A}^{FM}_{i,j} +\sum_i  \lambda_i  (\hat{n}_i -1 ),
\end{equation}
where we enforce the constraint with a site-dependent Lagrange multiplier $\lambda_i$ and the number operator for site $i$, given by $\hat{n}_i = \sum_m \hat{b}^\dagger_{i,m} \hat{b}_{i,m}$.  The $\hat{A}^{FM\dagger}_{i,j}$($\hat{A}^{FM}_{i,j}$) pair operators create(annihilate) a pair of bosons on sites $i$ and $j$ in total angular momentum state $|F,M\rangle$, and relate to the projection operators via $\hat{P}^F_{i,j}=\sum_{M} \hat{A}^{FM\dagger}_{i,j}\hat{A}^{FM}_{i,j}$.  Writing these pair operators in terms of the boson operators yields the relation $\hat{A}^{FM}_{i,j} = \sum_{m,n} C^{FM}_{m,n}  \hat{b}_{i,m} \hat{b}_{j,n}$, where the presence of the Clebsch-Gordan coefficients $C^{FM}_{m,n} = \langle f,m;f,n| F,M \rangle$ ensures that the pair operators rotate irreducibly as an object with angular momentum $F$.  In the form of Eq. (\ref{eq:FullHamiltonian}), we can write the Hamiltonian for a given atomic hyperfine spin $f$ by simply including the Clebsch-Gordan coefficients through the even-$F$ pairing channels, up to $2f$.  The straightforward calculation of these coefficients then provides for a much simpler study at large $f$.

Next, we mean field decouple the pairing operators $\hat{A}^{FM}_{i,j}$, first by expanding about the ground state expectation values $Q^{FM}_{i,j} = \langle \hat{A}^{FM}_{i,j} \rangle$, and then dropping terms of second order in the fluctuations $\delta \hat{A}^{FM}_{i,j}$.  This reduces the Hamiltonian to a quadratic form, given by
\begin{multline}\label{eq:MeanFieldHamiltonian}
\hat{H} = - \sum_{<i,j>} \sum_{F,M} \frac{1}{\bar{a}_F} \big(Q^{FM}_{i,j} \hat{A}^{FM \dagger}_{i,j} + Q^{FM*}_{i,j} \hat{A}^{FM}_{i,j}\\ - |Q^{FM}_{i,j}|^2 \big) +  \sum_i \lambda_i (\hat{n}_i -1 ),
\end{multline}
where henceforth $J =1$, and $\bar{a}_F = a_F/a_0$ denotes the scattering length of the $F$ channel relative to the singlet channel, $F=0$.  As our interest lies in translationally invariant states, we demand bond-independent mean fields, such that $Q^{FM}_{i,j}=Q^{FM}$.  The phase of the complex $Q^{FM}$ fields remains a U($1$) gauge freedom of the problem, and while gauge fluctuations may have important effects on spin liquid mean field theories,\cite{Wen1991} we do not consider them in the present approach.

For the investigation of spin liquid phases, as well as the study of atomic superconductors with non-trivial Cooper pairing,\cite{Ho1999} we note that decoupling the $\hat{A}^{FM}_{i,j}$ operators of Eq. (\ref{eq:FullHamiltonian}) proves more convenient than the single-mode approximation used in the study of spinor Bose-Einstein condensates.\cite{Kawaguchi2012}  In fact, our mean field Hamiltonian (\ref{eq:MeanFieldHamiltonian}) allows direct competition between exotic paired states described by the $Q^{FM}_{i,j}$ fields and the spinor BEC phases described by the boson field $\langle \hat{b}_{i,m} \rangle$.  In this way, our mean field theory may reproduce the results of the well studied spinor BEC mean field theories while also allowing for spin liquid ground states.  The outcome of this competition depends fundamentally on the strength of magnetic fluctuations, as a spin liquid state will only appear when sufficiently large spin fluctuations have melted the magnetic order of the spinor condensate phase.

\section{s-RVB ansatz}

To directly uncover a spin liquid phase in our model, Eq. (\ref{eq:MeanFieldHamiltonian}), we retain only the order parameter of the short-range resonating valence bond (s-RVB) state---namely, an isotropic nearest-neighbor pairing amplitude through the singlet channel.  In a pure s-RVB spin liquid state, all $F>0$ pairing channels have zero amplitude, and the spins exist in an equal superposition of nearest-neighbor singlets.  A ground state of this type preserves spin rotational and translational symmetry, making it one of the simplest spin liquid mean field theories possible for this model.  Furthermore, we may reach this s-RVB limit by taking $a_0 \to 0$ with $a_{F>0}$ fixed, since we then have $\bar{a}_{F}^{-1} \rightarrow 0$ for all $F>0$, and we see from Eq. (\ref{eq:MeanFieldHamiltonian}) that only the $Q^{00}$ pairing contributes.  The ability to access this limit by tuning a single parameter ($a_0$) may prove crucial to the eventual realization of this phase experimentally.

We employ the s-RVB ansatz explicitly in our formalism by substituting $\langle \hat{A}^{FM}_{i,j}\rangle = Q^{00} \delta_{F,0}$ in Eq. (\ref{eq:MeanFieldHamiltonian}), where the bond-independent complex number $Q^{00}$ represents the s-RVB order parameter.  Due to the equivalence of each site by symmetry we require a translationally invariant constraint, so that $\lambda_i = \lambda$.  Assuming periodic boundary conditions, we then exploit the lattice translational symmetry by Fourier transforming the bosons via $\hat{b}_{i,m} = \sum_{k} \hat{b}_{k,m} \mathrm{e}^{\mathrm{i} k \cdot R_i}/\sqrt{N}$, where the sum runs over all wavevectors $k$ in the first Brillouin Zone, $R_i$ denotes the Bravais lattice vector of site $i$, and $N$ gives the total number of sites in the lattice.  

Introducing the spinor $\Psi^T_{k,m} = ( \hat{b}_{k,m}, \hat{b}^\dagger_{-k,-m})$ allows us to compactly write the Fourier transformed Hamiltonian as
\begin{multline}\label{eq:s-RVBHamiltonian}
\hat{H} = \sum_{k,m} \Psi^\dagger_{k,m} h_{k,m} \Psi_{k,m} \\- \lambda N \frac{(2f+1)}{2}   + \frac{Z N |Q^{00}|^2}{2 a_0},
\end{multline}
where we have defined the $2 \times 2$ matrix
\begin{equation}\label{eq:2x2Hamiltonian}
h_{k,m} =
 \begin{pmatrix}
\lambda/2 & - \epsilon_k Q^{00} C^{00}_{m,-m} \\
- \epsilon_k Q^{00} C^{00}_{m,-m} & \lambda/2
 \end{pmatrix}.
\end{equation}
The lattice contribution to the Fourier transform yields $\epsilon_k = \sum_{<(i)j>}\mathrm{e}^{\mathrm{i} k \cdot (R_j - R_i)}$, while $<(i)j>$ denotes a sum over the $Z$ nearest neighbors $j$ of an arbitrary site $i$.  Throughout this paper we consider a 2-dimensional square lattice with a lattice spacing of unity, for which $Z=4$ and $\epsilon_k = 2 ( \mathrm{cos}[k_x] +  \mathrm{cos}[k_y])$.  Additionally, we have fixed the gauge by demanding a real $Q^{00}$ field.

The Clebsch-Gordan coefficients in Eq. (\ref{eq:2x2Hamiltonian}) play a crucial role in writing a $2 \times 2$-dimensional Hamiltonian for all spin $f$.  Primarily, the condition of the Clebsch-Gordan coefficients, that $C^{FM}_{m,n} = 0$ unless $m + n = M$, implies that the s-RVB state, which requires $M = 0$, retains only terms in which $n = -m$.  In other words, the matrix elements of our s-RVB Hamiltonian only ever connect a spin state $m$ with the corresponding state $-m$; a fact which leads to the chosen form of the spinor $\Psi^T_{k,m}$.  Compared to the general case of Eq. (\ref{eq:MeanFieldHamiltonian}), where we need a $2 (2f+1) \times 2 (2f+1)$-dimensional matrix at each $f$, the s-RVB ansatz  produces a dramatic mathematical simplification.

We now seek the ground state of the s-RVB Hamiltonian (\ref{eq:s-RVBHamiltonian}) in the presence of the one particle-per-site constraint.  Following the methods of Ref. \onlinecite{Sachdev1992}, we move to the basis of collective excitations $\hat{\gamma}_{k,\mu}$ by finding a linear transformation $M_{k,m}$ which diagonalizes the matrix $h_{k,m}$ while preserving the bosonic commutation relations, $[\hat{\gamma}_{k,\mu}, \hat{\gamma}^\dagger_{k',\mu'}] = \delta_{k,k'} \delta_{\mu,\mu'}$ and $[\hat{\gamma}_{k,\mu}, \hat{\gamma}_{k',\mu'}] = 0$.  After diagonalizing in this way, we find the collective excitations posses a dispersion given by $\omega_{k,m} = \sqrt{(\lambda/2)^2 - |\epsilon_k Q^{00} C^{00}_{m,-m}|^2}$ for the band corresponding to magnetic sublevel $m$, and we note that the $m$-independence of $|C^{00}_{m,-m}|=1/\sqrt{2f+1}$ forces complete degeneracy amongst these $2f+1$ bands.  Additionally, on a square lattice the dispersion takes a minimum value at $k = (0,0)$ and $k = (\pi,\pi)$, and the value of $\omega_{k,m}$ at these points defines the energy gap $\Delta =\sqrt{(\lambda/2)^2 - 16 | Q^{00}|^2/(2f+1)}$.  This gap will play a crucial role in the thermodynamic ground-state analysis to come.

\section{Results of the s-RVB ansatz}

To determine the ground state of the s-RVB Hamiltonian  in the thermodynamic limit ($N\to \infty$ with $N/V$ fixed) we solve the self-consistent equation, $Q^{00} = \langle \hat{A}^{00}_{i,j} \rangle$, in the presence of the constraint, $n_i = \langle \hat{b}^\dagger_{i,m} \hat{b}_{i,m} \rangle = 1$.  Writing the constraint in terms of the dispersion $\omega_{k,m}$ yields
\begin{equation}\label{eq:ConstraintEquation}
n_\gamma +  (2f+1) \int \frac{\mathrm{d}^2 k}{(2\pi)^2} \frac{1- \bar{\omega}_{k}^2}{2(\bar{\omega}_{k}^2+ \bar{\omega}_{k})} = 1,
\end{equation}
where we define $\bar{\omega}_{k} = 2 \omega_{k,m} / \lambda$ to clean up the notation a bit, and $n_\gamma$ denotes the condensate fraction of collective excitations, created by the $\hat{\gamma}_{k,\mu}$ operators, in states with energy $\Delta$.  With gapped excitations ($\Delta>0$), it costs finite energy to occupy these minimum energy states, and so the system prefers $n_\gamma = 0$.

\begin{figure}
\includegraphics[scale=0.22]{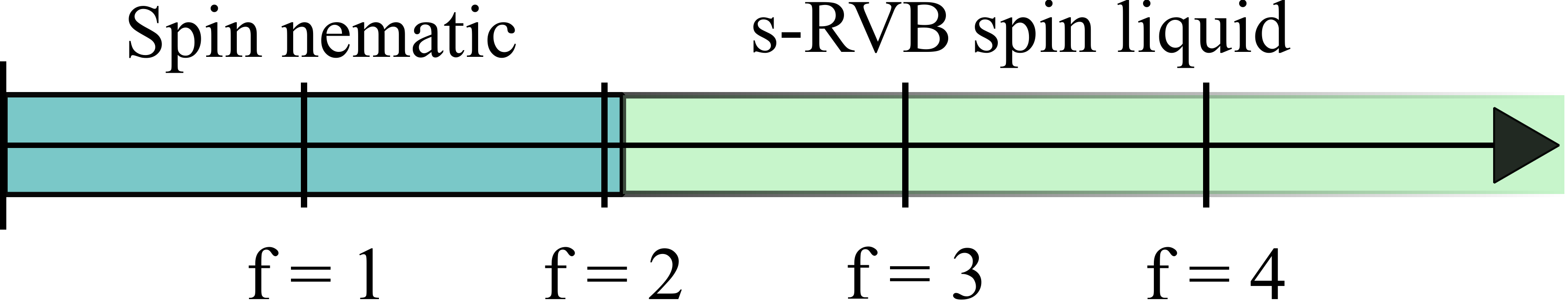}
\caption{\label{fig:Q00PhaseDiagram}Phase diagram of the s-RVB ansatz ($\alpha_{F>0} \to \infty$) as a function of $f$.  For $f\leq2$ the ground state is a spin nematic with $\langle \mathbf{S} \rangle = 0$ and $\langle S_x^2 \rangle = \langle S_y^2 \rangle \neq \langle S_z^2 \rangle$ on each site.  For $f>2$ however, degeneracy of the magnetic sublevels enhances fluctuations, and the ground state becomes a short-range resonating valence bond (s-RVB) spin liquid.}
\end{figure}

By numerically solving the constraint (\ref{eq:ConstraintEquation}) we find two scenarios shown in Fig. \ref{fig:Q00PhaseDiagram}.  The first occurs for $f\geq 3$, where one may satisfy the constraint with $\Delta > 0$.  The gapped excitations imply that the condensate fraction is zero, and so the $Q^{00}$ field characterizes the state completely, making it a realization of a pure s-RVB spin liquid.  On the other hand, for $f \leq 2$ one cannot satisfy the constraint with a gap, implying $\Delta = 0$ at $k = (0,0)$ and $k = (\pi , \pi)$.  The collective excitations condense at these points, and so $n_\gamma \neq 0$.  This restores the validity of the constraint, but one must now describe the state with a spinor of condensate parameters $\langle \hat{b}_{k,m} \rangle$, in addition to the $Q^{00}$ field.  A spinor of this type breaks spin rotational symmetry, implying a magnetically ordered ground state for $f \leq 2$.  At present, the spinor contains equally weighted $m$ and $-m$ pairs, and so possesses nematic symmetry with $\langle \hat{\mathbf{S}}_i \rangle = 0$ on each site.

The fact that for small spin we have a magnetically ordered spin nematic ground state, while for large spin we have a disordered spin liquid phase, results directly from the increasing number of magnetic sublevels as one moves to large $f$.  We understand this by noting that the integral in Eq. (\ref{eq:ConstraintEquation}),
\[
\langle \hat{n}_{i,m} \rangle = \int \frac{\mathrm{d}^2 k}{(2\pi)^2} \frac{1- \bar{\omega}_{k}^2}{2(\bar{\omega}_{k}^2+ \bar{\omega}_{k})} \leq 0.19\ldots,
\]
corresponds to the contribution from the non-condensed bosons of the $m$ band, $\langle n_{i,m}\rangle$, and has a maximum value $\simeq 0.19$ when $\Delta = 0$.  Degeneracy of the $2f+1$ bands then implies that for $\Delta >0$ we can write $\langle \hat{n}_i \rangle < 0.19*(2f+1) $, and so to satisfy $\langle \hat{n}_i \rangle = 1$ for finite $\Delta$, we must have $f>2$.

Again, we emphasize that these fluctuation driven states result directly from the increasing number of magnetic sublevels as one moves to larger spin.  The enlarged space through which the spins may interact enhances the fluctuations, melting magnetic order and driving the system into an s-RVB spin liquid phase.  Despite these positive results, to better describe the atomic species used in cold atom experiments we must study the more general problem, which allows scattering through the $F>0$ angular momentum channels.

\section{Nematic ansatz}

The most general case of the mean field Hamiltonian (\ref{eq:MeanFieldHamiltonian}) allows scattering through all total angular momentum channels, $F=0,2,\ldots,2f$.  However, with all $Q^{FM}$ fields allowed, the increasing size of the interaction space at large $f$ makes the Hamiltonian increasingly cumbersome to solve numerically.  So to efficiently probe the large-$f$ behavior as a function of the scattering lengths $a_F$, we retain only the order parameter of the spin nematic state, given by $\langle \hat{A}^{FM}_{i,j} \rangle = Q^{F0} \delta_{M,0}$.  With this set of mean fields each site will have $\langle \hat{\mathbf{S}} \rangle = 0$ and $\langle \hat{S}_x^2 \rangle = \langle \hat{S}_y^2 \rangle \neq \langle \hat{S}_z^2 \rangle$---the symmetry of a spin nematic.  Fortunately, since this retains only the $M=0$ pairing, we may again write the Hamiltonian with the 2-dimensional spinor $\Psi_{k,m}^T = (b_{k,m} , b_{-k,-m}^\dagger)$.  This again allows us to diagonalize in a straightforward manner, but we omit the details due to their similarity with the s-RVB case covered previously.

We motivate this ansatz by extending our results from the s-RVB ansatz, as shown in Fig. \ref{fig:Q00PhaseDiagram}, and by looking to the phase diagram for the spin 1, 2, and 3 spinor Bose-Einstein condensates.\cite{Ho1998,Ohmi1998,Imambekov2003,Puetter2008,Ciobanu2000,Ueda2002,Zhou2006,Diener2006,Santos2006,Suominen2007}  We find that our region of interest, parametrized by $a_F>0$ for all F and $a_{F'} \geq a_F$ for $F'>F$, lies entirely within the nematic sector of these phase diagrams.  Additionally, our ansatz consists of a linear combination of the uniaxial and biaxial spin nematic states, which are known to posses an accidental degeneracy at mean field level.\cite{Turner2007,Song2007}  Thus, it provides a suitable trial state for our specific parameter regime, capable of describing both a nematic spinor condensate and a $Q^{00}$-only s-RVB spin liquid phase.  We therefore proceed with this ansatz with the belief that our results describe the physically accessible states of the general $Q^{FM}$ model (\ref{eq:MeanFieldHamiltonian}).

While we have simplified the determination of the ground state for a general set of scattering lengths, a difficulty remains in how to best present the results graphically.  The number of scattering lengths grows as $f+1$, which on a phase diagram would require the introduction of an additional axis at each $f$.  To avoid this we seek an approximation which describes the various scattering lengths with a single parameter.  Guided by the s-RVB case, where we found that $a_0 \to 0$ favors the singlet pairing and induces an s-RVB spin liquid, we shall use the following approximation for the relative scattering lengths $\bar{a}_F = a_F/a_0$,
\begin{equation}\label{eq:aFApproximation}
\bar{a}_F = \begin{cases}
\alpha  &\text{for $F>0$}\\
1 &\text{for $F=0$}
\end{cases},
\end{equation}
where scattering lengths through non-zero angular momentum channels have equal magnitude, and differ from $a_0$ through the proportionality factor $\alpha$.  Varying $\alpha$ from $1$ to $\infty$ covers our original range of the scattering lengths---$a_F>0$ and $\bar{a}_F \geq 1$ for all $F$---while the $\alpha \to \infty$ limit recovers Eq. (\ref{eq:s-RVBHamiltonian}) directly.  While in real atomic systems the $\bar{a}_{F>0}$ are not generically equal, they effectively appear so when compared to $a_0$ in the $\alpha \to \infty$ limit, making this approximation especially useful for describing the spin liquid phase.  Most importantly, we may now construct a phase diagram as a function of $f$ and $\alpha$, since $\alpha$ is a parameter common to all spin $f$.

We note that applying Eq. (\ref{eq:aFApproximation}) takes us to an enhanced symmetry point of the original Hamiltonian (\ref{eq:FullHamiltonian})---namely, the bosonic analog to Wu's hidden symmetry found in large-spin Fermi gases.\cite{Wu2003,Wu2006}  Our results do not depend on this symmetry however, and we may show this by using the alternate approximation, $\bar{a}_F = \alpha' F + 1$, where the scattering lengths have a linear relationship with slope $\alpha'$.  This approximation does not generically posses symmetry higher than SU($2$), yet the results obtained coincide qualitatively with the results outlined in the next section using Eq. (\ref{eq:aFApproximation}).  The qualitative similarity stems from the fact that in each case, increasing $\alpha$ or $\alpha'$ effectively takes $a_0 \to 0$, and so the singlet pairing dominates---the crucial condition for obtaining a spin liquid phase in this model.

\section{Results of the nematic ansatz}

\begin{figure}
\includegraphics[scale=0.65]{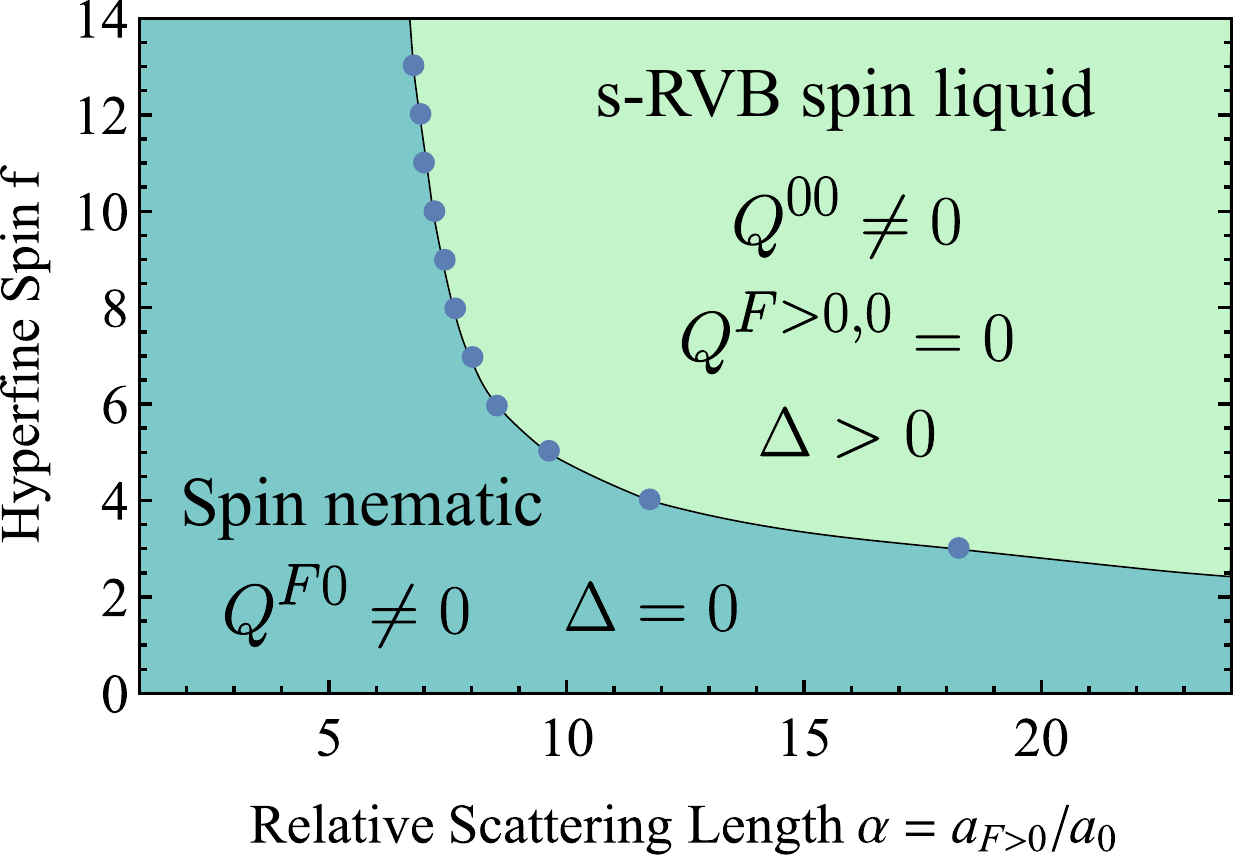}
\caption{\label{fig:QF0PhaseDiagram}Phase diagram of the nematic ansatz as a function of $f$ and $\alpha$, where $\alpha = a_{F>0}/a_0$ parameterizes the relative scattering lengths, as introduced by Eq. (\ref{eq:aFApproximation}).  The fluctuations responsible for the spin liquid state are enhanced by the increasing number of magnetic sublevels as one moves to large $f$, and by increased scattering through the singlet channel ($F=0$) as one moves to large $\alpha$.  For $f\leq 2$ the system is nematic for all $\alpha$, while for $\alpha = 1$ the spin liquid phase is not accessed at any $f$.}
\end{figure}

Figure \ref{fig:QF0PhaseDiagram} shows the phase diagram for $\alpha = [1,\infty)$ and $f = 1,2,\ldots,13$.  For $f\leq 2$ the system always form a nematic condensate, in agreement with our s-RVB solution in the $\alpha \to \infty$ limit.  On the other hand, for $f>2$ the system moves into the spin liquid phase for $\alpha$ greater than some critical value $\alpha_C$.  As we move to large $f$, we find that $\alpha_C$ decreases and the spin liquid region grows in size.  Again, decreasing $a_0$ relative to the other scattering lengths increases $\alpha$, and so by tuning a single parameter one may access the spin liquid phase for $f>2$ atoms.

We describe the behavior of the spin liquid region as follows.  Increasing $\alpha$ (or $\alpha'$) biases the system towards singlet pairing, which causes equal occupation of the Zeeman sublevels and maximizes magnetic fluctuations.  Additionally, moving to large $f$ increases the number of available magnetic sublevels, also enhancing fluctuations.\cite{Wu2010}  The shape of the spin liquid region as shown in Fig. \ref{fig:QF0PhaseDiagram} results from the cumulative effect of these two scenarios.  For $f \leq 2$, too few sublevels contribute to produce the necessary fluctuations, regardless of any biasing towards the singlet channel.  With $f>2$ but still small, the system requires strong biasing to reach the spin liquid phase.  While at large $f$, the multitude of participating sublevels allows access to the spin liquid phase with small biasing.  In light of this, large-spin atoms would require the least experimental tuning necessary to obtain the long sought-after spin liquid phase.

\section{Experimental Accessibility}

Upon inspection of the ``un-tuned'' scattering lengths, as conveniently compiled in Ref. \onlinecite{Kawaguchi2012} for the commonly used atoms---$^{87}$Rb and $^{23}$Na with $f = 1, 2$, and $^{52}$Cr with $f=3$---we see that $a_0$ and the smallest $a_{F>0}$ have roughly the same order of magnitude.  In each case, this places them near the $\alpha=1$ region of Fig. \ref{fig:QF0PhaseDiagram}, and predicts a spin nematic ground state in agreement with previous theoretical work.\cite{Ho1998,Ohmi1998,Imambekov2003,Puetter2008,Ciobanu2000,Ueda2002,Zhou2006,Diener2006,Santos2006,Suominen2007,Song2007}  However, upon tuning $a_0$ to small enough values via optical Feshbach resonance, a transition to the spin liquid phase may occur.  We note that this transition may even occur for $f \leq 2$ atoms as well, since fluctuations beyond mean field theory may actually enlarge the spin liquid region.

\begin{figure}
\includegraphics[scale=0.65]{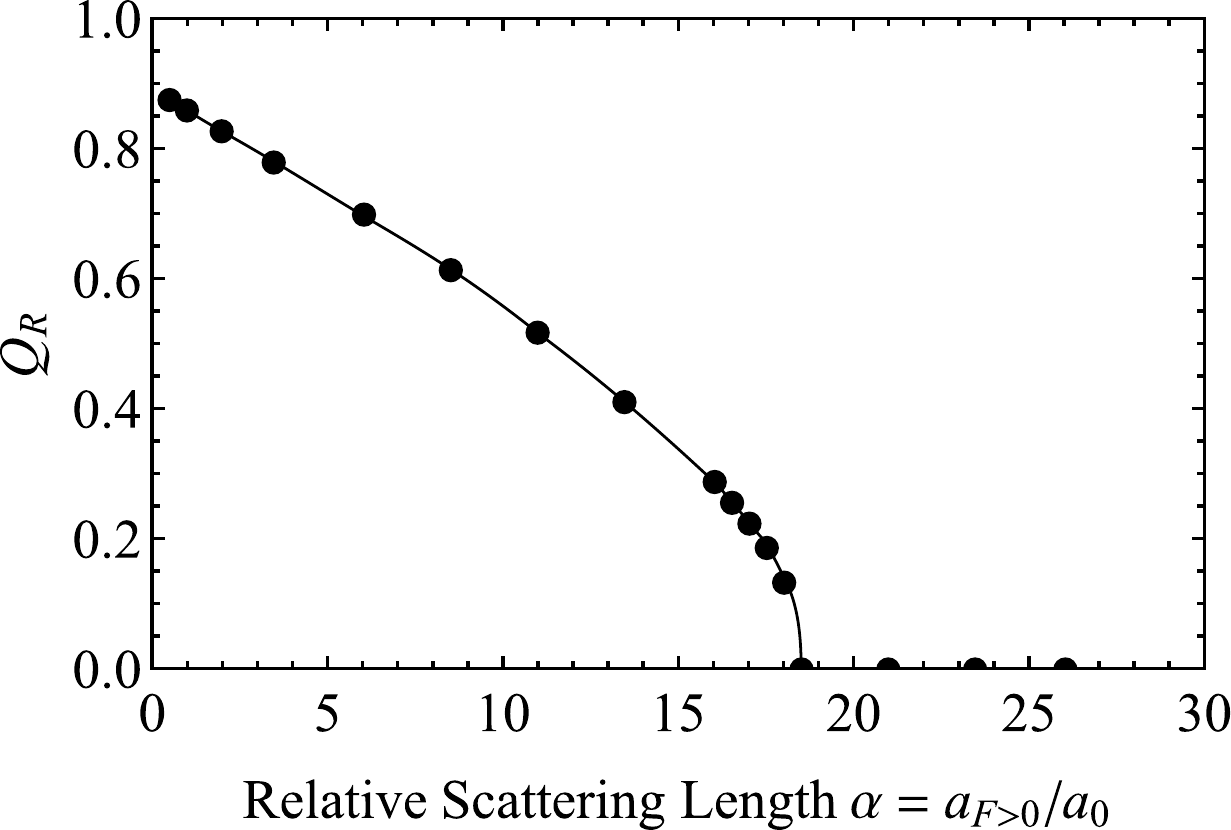}
\caption{\label{fig:OrderParameter}Relative contribution from the higher-order scattering channels ($F>0$) for an $f=3$ system, as captured by the $Q_R$ parameter defined in Eq. (\ref{eq:OrderParameter}).  In the spin liquid phase ($\alpha \gtrsim18$) all pairings except the singlet pairing $Q^{00}$ are negligible, while in the spin nematic phase ($\alpha \lesssim 18$) pairing through the non-zero angular momentum channels becomes relevant.  Measurement of this parameter could distinguish between the phases.}
\end{figure}

The relative contribution from the $F>0$ pairing channels, 
\begin{equation}\label{eq:OrderParameter}
Q_R=\frac{\sum_{F>0} |Q^{F0}|^2}{\sum_{\text{all }F}|Q^{F0}|^2},
\end{equation}
represents a potential order parameter for the spin liquid-to-spin nematic phase transition.  Fig. \ref{fig:OrderParameter} shows the behavior of this quantity for an $f=3$ system when tuned across $\alpha$.  This quantity is similar to the singlet-fraction measured in Ref. \onlinecite{Trotzky2010}, and may allow the observation of a spin liquid phase experimentally.  Additionally, by spatially resolving vortices in the $Q^{FM}_{i,j}$ fields via photoassociation intensity experiments\cite{Abraham1996,Prodan2003,Junker2008,Dutta2013,Dutta2014} one may investigate vison excitations in the system, in a similar manner to the ``vison experiment'' conducted by Kam Moler and collaborators for high-$T_C$ cuprate superconductors.\cite{Bonn2001}  Overall, the increasingly varied techniques used in the preparation and characterization of cold atomic systems may provide several avenues for the eventual observation of these novel states.

There remains two primary challenges in the quest to observe large-$f$ spin liquids, however.  Firstly, anisotropic dipole-dipole interactions---which grow with the spin as $f^2$---may obstruct the investigation at large spin.  For example, several recently trapped isotopes of Dysprosium have an extremely large spin of $f = 7$ and $8$, and the dipole-dipole interactions in these systems are believed to affect the ground-state physics in a non-perturbative way.\cite{Tang2015}  Essentially, the dipole-dipole energy sets a lower bound on the allowed tuning of $a_0$, below which our approximation breaks down and we must account for these interactions explicitly.\cite{Lahaye2009}  Fortunately, for $f=3$ Chromium, the dipole-dipole interactions do not significantly affect the ground-state physics,\cite{Ho1998} allowing use of our mean field description (\ref{eq:MeanFieldHamiltonian}).  The remaining challenge is then to cool the gas sufficiently for the observation of magnetic exchange, since the entropy obtained initially by the gas determines the entropy present after adiabatically ramping into a Mott state.\cite{Koetsier2008}  Fortunately, large-spin atoms carry away more energy during evaporative cooling due to their large manifold of hyperfine sublevels, in a similar manner to the SU($N$) alkali systems at large $N$,\cite{Cazalilla2014} which may actually facilitate cooling in these large-spin systems.

When the experimental challenges have been overcome, Mott insulating ultracold bosonic systems could provide a rich environment in which to observe and study the long sought-after quantum spin liquid phase.  In these systems, the unique mechanism of whole-atom exchange allows one to increase fluctuations by simply using atoms with a larger hyperfine spin; in combination with the tunability of the interactions via optical Feshbach resonance, this makes the realization of an atomic spin liquid a realistic prospect.  Given the richness of the large-f spin models, there may even be a whole class of atomic spin liquid-like phases, each with different spin, lattice geometry, dimensionality, and interaction range.  Fortunately, as the experimental control and manipulation of large-spin atoms improves, we only edge closer to the exciting time when we may capture the elusive spin liquid phase in the novel environment of cold atoms.

\emph{Acknowledgments---} We thank Arun Paramekanti and Erich Mueller for useful discussions.

\end{document}